\RequirePackage{lineno}
 \documentclass[final,5p,times,twocolumn]{elsarticle}

\usepackage{epsfig}

\usepackage{amssymb}
\usepackage{amsmath}
\usepackage{mathtools}
\usepackage{extarrows}
\usepackage{wasysym}
\usepackage{breqn}
\usepackage{amscd}
\usepackage{feynmf}

\begin{document}

\begin{frontmatter}



\title{On the emergence of natural singularities and state transitions in living patterns}


\author{Akos Dobay}

\address{Institute of Evolutionary Biology and Environmental Studies, Winterthurerstrasse 190, University of Zurich, CH-8057 Zurich, Switzerland}

\begin{abstract}
As far as human perceptions and rational thinking are concerned, contradictions constitute a non negligible part of our reality. We often refer to these phenomena, in a more informal way, as the \textit{chicken or the egg} causality dilemma. However, it is not clear whether the \textit{chicken or the egg} dilemma exists only within the scope of our perceptions, or contradictions have a deeper meaning towards our understanding of reality. Here we argue that if there is an element of reality such that can be adequately described in terms of the \textit{chicken or the egg} dilemma, then it might lead to a spontaneous symmetry breaking by creating an alternate entity, capable of ultimately separating the chicken from the egg. We propose a formalism to describe such mechanism and discuss how it can be applied to phenomena to describe the natural emergence of singularities and state transitions in living systems. 
\end{abstract}


\end{frontmatter}


\section{Introduction}
\label{introduction}
What comes first: the chicken or the egg? Everyone knows it cannot be the egg since the chicken has to lay it first. Conversely, it cannot be the chicken since it comes out of an egg. This problem, known as the \textit{chicken or the egg} dilemma, has been used to describe paradoxical situations. Paradoxes are well known and are of great interest, in the scientific community and in mathematics, because they represent a challenge to any form of rational thinking \cite{Hofstadter1979}. Paradoxes have fascinated generations of philosophers for the same reason. The formulation of the \textit{chicken or the egg} dilemma has also inspired several artists. One of the most popular is certainly the Dutch artist M. C. Escher (1898 -- 1972). An example is his famous lithograph called \textit{Drawing hands} (1948), in which he depicts two hands, facing each other on a sheet of paper in the paradoxical act of drawing one another into existence.

A paradoxical situation or a \textit{chicken or the egg} dilemma appears in many different fields of knowledge.  Whether it is in mathematics, if we think of G\"odel's incompleteness theorems \cite{Godel1931}, in quantum mechanics and the so-called EPR paradox \cite{Einstein1935,Everett1957}, or even in biology when it comes to find out how replication started during the early stages of life \cite{Eigen1971}, there is always, at least in appearance, a \textit{chicken or the egg} dilemma. The use of the words ``in appearance'' in the previous sentence conveys an important aspect of the problem we would like analyze with the present formalism. At the scientific level, the problem can be formulated as follow: ``Is there a mechanism behind the resolution of an internal contradiction within a given physical system?'' In our case the concept of a physical system  corresponds to a biological organism which has to adapt and might be subjected to natural selection. The answer to this question will determine whether all the physical constituents of an internal contradiction can be unambiguously identified and translated into an algorithm or the resolution happens in a dimension for which no tangible constituents exist.

Among humans, paradoxical situations occurs in the course of time when the informational content of a system is such that it cannot be resolved rationally. Hence, when facing a paradox, a decision-maker might be confronted with the dilemma of not being able to make an appropriate choice within a reasonable amount of time. Moreover, when the decision has to be taken under environmental pressure, it naturally creates tension, whose intensity varies among individuals. This tension will sooner or later induce a decision, breaking the paradox into two distinct solutions. But either of these solutions is absolute: they rely on each other. Paradoxes also lie at the core of fundamental theories \cite{Hofstadter1979}. In our attempt to find absolute criteria upon which to build a satisfactory rational reference frame, independent of any particular perceptions of the world and free from contradictions, the \textit{chicken or the egg} dilemma stands as a true paradigm.

Can we answer our question  or circumvent its limitations if no answer can be provided? Answering the question turns out to be still out of the scope of our present formalism. However we argue that its limitations can be at least weakened using an appropriate thought pattern. To achieve our goal, we rely on the working hypothesis that an internal contradiction requires an additional dimension, apart from those that constitute the paradox parameter space, to be resolved. However, there might be no appropriate mathematics to characterize it; consequently, no calculations are possible. If the result of such process is transposed back onto its original parameter space, it might be possible to develop a measure to quantify its effects.

Hence, the \textit{chicken or the egg} dilemma becomes a paradigm where the question does not necessarily require the answer to which one came first, and to subsequently establish a logical order. Rather , it is a paradigm for situations where the two elements can co-exist, but accompanied by a tension that forces the contradiction to be resolved into measurable elements such as space and time \cite{Dobay2008}. What appears to be a limitation in our rational thinking or our perception of the world can be instrumental in understanding phenomena. This paradigm may explain the emergence of individuals or elements capable of mirroring their environment, whenever paradoxical situations arise where the available information is not sufficient to drive a state transition, but where external constraints are necessitating the transition to occur nonetheless: effectively, a decision-making process \cite{Dobay2008}. It might also serve as a hypothetical explanation for the emergence of particle-antiparticle pairs from vacuum breakdown \cite{Schwinger1951}, in case the boson is a form of tension traveling though space and time until it materializes into a pair.

The class concept is useful for describing similar entities. Formally, a class is a collection of objects that unambiguously share the same properties. Similarly, in object-oriented programming, the notion of class is an abstraction that permits the encapsulation of a collection of properties common to objects belonging to it. By considering the \textit{chicken or the egg} dilemma as a class of related problems, we can formulate the paradigm in more general terms and propose a specific formalism to test the underlying mechanism behind such phenomena, and their implications on our understanding of nature. Moreover, any system, which has the ability to processt information from its environment, can potentially take decisions. Commonly, this ability is attributed to higher order organisms capable of cognition, but not to other systems such as viruses or bacteria, although some authors are likely to investigate this possibility \cite{Xie2010, Weitz2008}. It is even harder to conceive that this ability can be transposed to purely physical systems. For this essay, we assume that reflection is a natural phenomenon that emerges when a physical system -- living or not -- composed of a combinatorial set of building blocks reaches the critical situation where environmental pressure forces the system to undergo a morphological or a topological transition even if temporal delays in information processing hinder a logical decision from being made. Information lags result in the superposition of the past and present states, creating an internal contradiction. From such a starting set of combinatorial building blocks, any structure from a single atom to an entire galaxy may run into a paradoxical situation.

\section{Description of the formalism}
\label{formalism}
As we mentioned in the introduction, the present formalism presents the \textit{chicken or the egg} dilemma  as a class of related problems $C_{\alpha \beta}$. Therefore, any element of $C_{\alpha \beta}$ is characterized by an internal contradiction that transitions from a dimension $d_{1}$ into a measurable space $d_{2}$. The transition from  $d_{1}$ to $d_{2}$ requires the emergence of a singularity that we formally indicate as $\mathcal{T}$. Here, we show how to quantify an element of $C_{\alpha \beta}$.

The \textit{chicken or the egg} dilemma can be represented as the superposition of two mutually exclusive entities $\alpha$ and $\beta$ in the non-spatial dimension $d_{1}$:

\begin{equation}
\begin{pmatrix} \alpha \\ \beta \end{pmatrix}_{d_{1}}.
\label{equation1}
\end{equation}
\noindent
Once the internal contradiction has been resolved in any type of space $d_{2}$ with a well-defined measure, then (\ref{equation1}) becomes

\begin{equation}
\begin{pmatrix} \alpha & \beta \end{pmatrix}_{d_{2}}.
\label{equation2}
\end{equation}
\noindent
 Hence, at the very basic level we have

\begin{equation}
\begin{pmatrix} \alpha \\ \beta \end{pmatrix}_{d_{1}}  \xlongleftrightarrow{\mathcal{T}}  \begin{pmatrix} \alpha & \beta \end{pmatrix}_{d_{2}}
\label{equation3}
\end{equation}
\noindent
where $\mathcal{T}$ results from the tension created in $d_{1}$ by the internal contradiction. The relation given by (\ref{equation3}) is a representation of an object of $C_{\alpha \beta}$, basically the atomic structure of an internal contradiction. The bidirectional arrow between the two members of (\ref{equation3}) means that we make no assumptions regarding the origin of internal contradictions. Now that we have given a formal representation of a particular element of $C_{\alpha \beta}$ we need to provide a mathematical toolset able to quantify it. As in Bayesian inference method \cite{Bayes1763}, our approach relies on the concept of probability as a measure of uncertainty, as partly introduced in \cite{Dobay2008}.

Briefly, to describe the relation between the information content of the system and a transition, we characterize the state of a system by its confinement in a parametric space, i.e. by its contour $c$. In other words, behind each parameter of a system, we assume there is a physical agent $(a_{i})$, such that the existence of a contour can be represented by a set of parameters $(p_i)_{i = 1, 2, \dots , n}$ and their specific values $\textbf{x}=(x_i)_{i = 1, 2, \dots , n}$, leading to a given morphology or topology that defines the state; these parametric values belong to a statistical ensemble. Therefore, if we assume that the specific values are variables which have a probabilistic distribution, then the contour $c(x_i)_{i = 1, 2, \dots , n}$ can be defined as a distribution function of the set of parameters and their respective variable $x_{i}$. We further assume that $f(\textbf{x})$ represents the joint probability density function associated with $c$. We also assume that the ranges within which the specific values of the parameters while variable are still representative of the state, and lie in the space of real values $\Omega = \prod_{i=1}^{n}[a_{i},b_{i}]$ which is included in $c(x_i)_{i = 1, 2, \dots , n}$. Thus, the probability of finding the parameters $(p_i)_{i = 1, 2, \dots , n}$ inside a contour can be calculated by integrating the joint probability density function over all these intervals
\begin{equation}
\textnormal{Pr} \biggl ( c(x_i)_{i=1,2,\ldots,n} \biggr)=\int_{a_1}^{b_1} \int_{a_2}^{b_2} \ldots \int_{a_n}^{b_n} f(\textbf{x}) \: d\textbf{x}.
\label{equation4}
\end{equation}
In addition, the probability associated with the parameter $p_{i}$ alone can be deduced from equation \ref{equation4} by integrating the joint probability density function over all values of the $n-1$ other parameters
\begin{equation}
f_{\textbf{x}}(x_i)=\int f(\textbf{x}) \: dx_1 \: \ldots \: dx_{i-1} \: dx_{i+1} \: \ldots \: dx_n.
\label{equation5}
\end{equation}
\noindent
From equation \ref{equation5}, one can finally calculate the probability associated with the parameter $p_{i}$ at a given point in time

\begin{equation}
\textnormal{Pr} \biggl ( p_{i,t}^{c(x_i)_{i=1,2,\ldots,n}} \biggr)=\int_{a_i}^{b_i} f_{\textbf{x}}(x_{i,t}) \: dx_i.
\label{equation6}
\end{equation}

During a transition, the system abandons its old contour and adopts a new one. An orderly transition is characterized by the absence
of an overlap between the two contours, i.e. the parametric values of the old and new states are variable within non-overlapping intervals. Now let
$p_{i}$, be the $i$-th parametric value at a given time point, then the difference $d_{t}(c_{1},c_{2}) \in [-1,1]$ between two contours $c_{1}$ and $c_{2}$, can be defined as
\begin{equation}
d_{t}(c_1,c_2) \mathrel{\mathop:}= \frac{1}{n} \sum_{i=1}^{n} \textnormal{Pr} \biggl ( p_{i,t}^{c_2} \biggr)-\textnormal{Pr} \biggl ( p_{i,t}^{c_1} \biggr)
\label{equation7}
\end{equation}
\noindent
wherein $\textnormal{Pr}\biggl (p_{i,t}^{c_{j}}\biggr)$ stands for the probability of the $i$-th parameter being in the domain of the $j$-th contour. The value of $d_{t}(c_{1},c_{2})$ indicates at a given time point how many agents contribute to each contour $c_{1}$ and $c_{2}$. Another way to define $d_{t}(c_{1},c_{2})$ would have been by the Kullback-Leibler divergence \cite{Kullback1951}. The Kullback-Leibler divergence is formally given by

\begin{equation}
D_{KL}(f_{c_{1}} \: || \: f_{c_{2}}) = \int_{-\infty}^{\infty} \ln \biggl ( \frac{f_{c_{1}}(x)}{f_{c_{2}}(x)} \biggr) f_{c_{1}}(x) \: dx
\label{equation8}
\end{equation}
\noindent
and expresses the difference between two probability distributions $f_{c_{1}}$ and $f_{c_{2}}$ when the distribution $f_{c_{1}}$ represents the reference model against which one quantifies the loss of information when using the model represented by the distribution $f_{c_{2}}$. However, the contours $c_{1}$ and $c_{2}$ are not necessarily comparable -- and most of the time they are not. This brings us to introduce a simple distance function.

We can now define the strain $S_{t}(c_1,c_2) \in [0,1]$ associated with the internal contradiction as the inverse of the absolute value of $d_{t}(c_1,c_2)$

\begin{equation}
S_{t}(c_1,c_2) \mathrel{\mathop:}= \frac{1}{\mid d_t (c_1,c_2) \mid}.
\label{equation9}
\end{equation}
Here we assume $S_{t}(c_1,c_2)$ to be strictly positive. To the notation defined by (\ref{equation2}) can now correspond a probabilistic distribution measure such that
\begin{equation}
\begin{pmatrix} c_{1} & c_{2} \end{pmatrix}_{d_{2}} \mathrel{\mathop:}= d_t (c_1,c_2).
\label{equation10}
\end{equation}
\noindent
Similarly the internal contradiction represented by  (\ref{equation1}) can be written as
\begin{equation}
\begin{pmatrix} c_{1} \\ c_{2} \end{pmatrix}_{d_{1}} \mathrel{\mathop:}= S_{t}(c_1,c_2).
\label{equation11}
\end{equation}
What we described as an emerging singularity $\mathcal{T}$ can now be quantified by its dual manifestation in $d_{1}$ and $d_{2}$.
\section{Discusion}
\label{discussion}
The physicist Erwin Schr\"odinger wrote that physical models do not represent reality. They are simply useful when they are adequate for describing we observe \cite{Schrodinger1952}. The present formalism has multiple motivations. First, it reconnects with the very basic desire to find general design principles that could explain the features of underlying phenomena. Second, it represents an attempt to rationally model what we experience. Using Schr\"odinger's terms, the formalism we propose aims to adequately describe what we observe and connect different phenomena.

Many authors, whether they were scientists, philosophers, poets or writers came to a similar conclusion: a fundamental uncertainty, or ambiguity, prevents us from accessing a complete picture of the reality within a single rational framework \cite{Godel1931,Einstein1935,Heisenberg1927}. It seems that at the core of every natural system lies a fundamental contradiction, and this observation encompasses rational thinking. Like in philosophy, one can pose the non-existence of God as a postulate and draw the ethical consequences \cite{Sartre1943}, one can also assume the existence of an internal contradiction in every natural system and explore its ramifications.

In this respect, the morphological distance $d_{t}(c_1,c_2)$ and the tension $S_{t}(c_1,c_2)$ associated with it constitute a very general approach for quantifying multifactorial situations whose outcome is in an internal contradiction. A good example of such a situation are transitions in single cell organism.

Cellular mechanisms are often subjected to transitions; in eucaryotes, transitions during cell division is orchestrated into ordered states known as the cell cycle. The cycle is flanked by two main checkpoints between the passage from $G_1$ to $S$ and the passage from $G_2$ to $M$. $G_1$ is the state prior to genome duplication or DNA synthesis ($S$). $G_2$ is the state prior to mitosis ($M$) or cell division. The transition from one state to another is most likely based on feedback loops and molecular signaling, where cycle-specific molecules and proteins are synthesized and degraded based on extracellular inputs. In the case of the cell cycle, cyclin-dependent kinases (Cdks) are known to play an important role \cite{Murray2004}. Conflicting or incomplete information is not unusual in this process, leading to unresolved situations \cite{Mankouri2013}. Hence, being able to measure the level of each contributing signals permits the computation of $d_{t}(G_1,S)$ and $S_{t}(G_1,S)$, or $d_{t}(G_2,M)$ and $S_{t}(G_2,M)$. The problem of having conflicting or incomplete information can also result from an incomplete screening of the environment. In this case, a contradiction can be minimized by a more appropriate distribution of cellular receptors or with a different set of pathways better suited to analyze the situation. Whether the cell can fully resolved the space- and time-dependent orchestration of its own division is still, in our opinion, an open question.

Another concrete example can be drawn from the field of epigenetics. Bivalent lineage-specific genes constitute a large family composed of more than 2,000 members  \cite{Bernstein2006, Barski2007, Cui2009, DeGobbi2011}. They are characterized by the presence of both activating and repressive chromatin marks, H3K4me3 and H3K27me3. Most bivalent domains are often associated with the loci of key developmental transcription factors, morphogens, as well as cell surface molecules \cite{Bernstein2006, Mikkelsen2007}. A prevailing hypothesis about bivalent genes is that H3K27me3 suppresses the expression of lineage control genes, but the H3K4me3 keeps it in a poised state for activation \cite{Bernstein2006}, although the idea of epigenetic predetermination of such poised states was recently challenged in \cite{DeGobbi2011}. In thist study, the authors emphasize that the levels of H3K4me3 and H3K27me3 associated with bivalent domains are widely variable; together with RNA polymerase occupancy data and transcript information, they hypothesize that H3K4me3 is a marker for variable degrees of stochastic transcription, while H3K27me3 serves to regulate the expression of bivalent genes. Nonetheless, the incidences of stochastic transcription reported were at basal level, and the bivalent genes could still be considered poised \cite{DeGobbi2011}. On differentiation, most of the bivalent modifications are resolved into either activated or repressed states \cite{Bernstein2006}. It is consequently plausible to hypothesize that ambivalence at the level of gene expression corresponds to cases where two, or possibly, more gene regulation outcomes are linked in a shared state which is undefined until commitment. The final outcome can be modeled in terms of the distance $d_{t}(H3K4me3,H3K27me3)$ and the tension $S_{t}(H3K4me3,H3K27me3$) depending on the level of histone methylation and how many transcription factors are effectively present to trigger the state transition.

In a similar way, with the rise of virtual social networks, it is possible to have access to a very large set of indexes about what individuals in a society feel at a given time point. Translating opinions into variables, one could technically calculate their probability density function to estimate the social tension in a society.

To further strengthen the essence of our formalism, the concept of cooperation among individuals provides a straightforward example. Individuals like humans or mammals can either cooperate or compete. If we assume that competition is the dual aspect of cooperation (i.e. that cooperation cannot exist without competition), what we consider as competition or cooperation is only a matter of where we set the limit. One can increase competition and lower cooperation or the other way around. What makes competition really effective is the ability of some entities to find a way to take advantage of other entities. In this manner, the tension of the internal contradiction is broken, resulting in an asymmetry in space and time. This asymmetry can be amplified to the point that the duality becomes residual for an external observer.

We also want to emphasize that in our opinion, neither $d_{t}(c_1,c_2)$ nor $S_{t}(c_1,c_2)$ reach extrema, which are $\{-1,1\}$ and $\{0,1\}$ respectively. Indeed, a well-defined morphology always contains a fraction of its parameter values outside what intrinsically constitutes its manifestation. Consequently, the tension can never be totally absent from the system.

Our formalism also constitutes an interesting framework to discuss randomness. Randomness is assumed to be consubstantial with several natural phenomena. The disintegration of particles is a typical example of process in which events occur in a random manner. Brownian motion of protein complexes within cell nucleus is another example \cite{Siebrasse2008}. Predictions can be made only based on probabilities. Similarly, the mathematical development realized for building random number generators turns out to be a non-trivial task, and only finite series can be achieved. Such series are usually called pseudo-random number generators (PRNGs). How random are transcendental numbers such as $\pi$ or $e$ is still an open question \cite{Pincus1997}.  Here we argue that a true paradoxical situation can only generate a random outcome and constitutes a natural source of randomness in living processes, which in turn may act as one of the evolutionary forces in nature. In that case, a clear venue where our formalism may have interesting applications is in artificial intelligence. Being able to realize a device based on an intrinsic tension, and capable of producing its own decisions, via a singularity $\mathcal{T}$, can lead to a new generation of self-evolving entities.

Finally, one can point out that we have not investigated multiple morphologies, where the outcome relies on multiple contours. Having to choose between multiple possibilities clearly constitutes the most common situations in nature. However once the possibilities have been weighted for a subsequent action, all these possibilities are narrowed down to at most two possibilities; and this is clearly what constitutes the difficulty of making a choice.

\section{Conclusion}
\label{conclusion}
When the Copenhagen interpretation of quantum mechanics was formulated by physicists such as Niels Bohr, Werner Heisenberg and Erwin Schr\"odinger among others  \cite{Rechenberg1982}, concepts like entanglement, wave function and uncertainty undermined the idea of a complete picture of a physical reality. Even if these concepts sounded new to scientists used to work with classical mechanics, it is clear that any attempt to fully describe reality in a deterministic way would have sooner or later failed. Who can possibly predict what will happen tomorrow? Possibilities in the future are entangled and even after they happened, there is not always an objective measure to decide which of these possibilities really occurred.

The present formalism suggests the central perspective of internal contradiction to examine phenomena; this might resemble to some extent the concept of entanglement in quantum mechanics, since the intrinsic ambiguity contained in the wave function of a multiple states system is resolved upon measurement. In contrast to the formalism developed in quantum mechanics, the physical manifestation of the internal contradiction described by the the \textit{chicken or the egg} dilemma does not lead to a complete loss of correlation between the contours. As we already mentioned, an internal contradiction is always characterized by a tension and a difference. Hence, when a pattern emerges from an internal contradiction, it has to minimized its tension into one of the contours. In other words, the pattern does not have an independent reality. A pattern emerges when the circumstances allow the reduction of tension towards one of the contours, but the ambiguity remains as manifestation of tension.

The formalism we proposed to conceptualize an observation is believed to be essential in many different phenomena -- that is, a scale-free design principle, whose applications are not necessarily limited to biological systems.

\section{Acknowledgment}
\label{acknowledgment}
The author thanks Homayoun C. Bagheri for the fruitful discussions and his support during the writing of the manuscript and Maria Pamela Dobay for her valuable comments and proofreading.




\bibliographystyle{elsarticle-num}
\bibliography{bibliography.bib}

\begin{thebibliography}{10}
\expandafter\ifx\csname url\endcsname\relax
  \def\url#1{\texttt{#1}}\fi
\expandafter\ifx\csname urlprefix\endcsname\relax\def\urlprefix{URL }\fi
\expandafter\ifx\csname href\endcsname\relax
  \def\href#1#2{#2} \def\path#1{#1}\fi

\bibitem{Hofstadter1979}
D.~R. Hofstadter, G{\"o}del, Escher, Bach : an eternal golden braid, Basic
  Books, New York, 1979.

\bibitem{Godel1931}
K.~G{\"o}del, {\"U}ber formal unentscheidbare s{\"a}tze der principia
  mathematica und verwandter systeme, I. Monatshefte f{\"u}r Mathematik und
  Physik 38 (1931) 173--198.

\bibitem{Einstein1935}
A.~Einstein, B.~Podolsky, N.~Rosen, Can quantum-mechanical description of
  physical reality be considered complete?, Phys. Rev. 47 (1935) 777--780.

\bibitem{Everett1957}
H.~Everett, ``relative state'' formulation of quantum mechanics, Rev. Mod.
  Phys. 29 (1957) 454--462.

\bibitem{Eigen1971}
M.~Eigen, Selforganization of matter and the evolution of biological
  macromolecules, Naturwissenschaften 58~(10) (1971) 465--523.

\bibitem{Dobay2008}
A.~Dobay, W.~van Toledo, Informational processing of morphological and
  topological transitions in biology, Arkhai 13 (2008) 5--16.

\bibitem{Schwinger1951}
J.~Schwinger, On gauge invariance and vacuum polarization, Phys. Rev. 82 (1951)
  664--679.

\bibitem{Xie2010}
Z.~Xie, L.~E. Ulrich, I.~B. Zhulin, G.~Alexandre, Pas domain containing
  chemoreceptor couples dynamic changes in metabolism with chemotaxis,
  Proceedings of the National Academy of Sciences 107~(5) (2010) 2235--2240.

\bibitem{Weitz2008}
J.~S. Weitz, Y.~Mileyko, R.~I. Joh, E.~O. Voit, Collective decision making in
  bacterial viruses, Biophysical Journal 95~(6) (2008) 2673--2680.

\bibitem{Bayes1763}
M.~Bayes, M.~Price, An essay towards solving a problem in the doctrine of
  chances. by the late rev. mr. bayes, f. r. s. communicated by mr. price, in a
  letter to john canton, a. m. f. r. s., Philosophical Transactions 53 (1763)
  370--418.

\bibitem{Kullback1951}
S.~Kullback, R.~A. Leibler, On information and sufficiency, Annals of
  Mathematical Statistics 22~(1) (1951) 79--86.

\bibitem{Schrodinger1952}
E.~Schr{\"o}dinger, Science and Humanism: Physics in Our Time, Cambridge univ.
  Press, New York, 1952.

\bibitem{Heisenberg1927}
W.~Heisenberg, {\"U}ber den anschaulichen inhalt der quantentheoretischen
  kinematik und mechanik, Zeitschrift f{\"u}r Physik 43~(3-4) (1927) 172--198.

\bibitem{Sartre1943}
J.-P. Sartre, L'{\^E}tre et le N{\'e}ant, \'Editions Gallimard, Paris, 1943.

\bibitem{Murray2004}
W.~A. Murray, Recycling the cell cycle: Cyclins revisited, Cell 116~(2) (2004)
  221--234.

\bibitem{Mankouri2013}
H.~W. Mankouri, D.~Huttner, I.~D. Hickson, How unfinished business from s-phase
  affects mitosis and beyond, The EMBO Journal 32~(20) (2013) 2661--2671.

\bibitem{Bernstein2006}
B.~E. Bernstein, T.~S. Mikkelsen, X.~Xie, M.~Kamal, D.~J. Huebert, J.~Cuff,
  B.~Fry, A.~Meissner, M.~Wernig, K.~Plath, R.~Jaenisch, A.~Wagschal, R.~Feil,
  S.~L. Schreiber, E.~S. Lander, A bivalent chromatin structure marks key
  developmental genes in embryonic stem cells, Cell 125~(2) (2006) 315--26.

\bibitem{Barski2007}
A.~Barski, S.~Cuddapah, K.~Cui, T.~Roh, D.~Schones, High-resolution profiling
  of histone methylations in the human genome, Cell 129 (2007) 823--37.

\bibitem{Cui2009}
K.~Cui, C.~Zang, T.~Roh, D.~Schones, R.~Childs, W.~Peng, K.~Zhao, Chromatin
  signatures in multipotent human hematopoietic stem cells indicate the fate of
  bivalent genes during differentiation, Cell Stem Cell 4~(1) (2009) 80--93.

\bibitem{DeGobbi2011}
M.~D. Gobbi, D.~Garrick, M.~Lynch, D.~Vernimmen, J.~R. Hughes, N.~Goardon,
  S.~Luc, K.~M. Lower, J.~A. Sloane-Stanley, C.~Pina, S.~Soneji, R.~Renella,
  T.~Enver, S.~Taylor, S.~E.~W. Jacobsen, P.~Vyas, R.~J. Gibbons, D.~R. Higgs,
  Generation of bivalent chromatin domains during cell fate decisions,
  Epigenetics {\&} Chromatin 2010 3:2 4~(1) (2011) 9.

\bibitem{Mikkelsen2007}
T.~Mikkelsen, M.~Ku, D.~Jaffe, B.~Issac, E.~Lieberman, G.~Giannoukos,
  P.~Alvarez, W.~Brockman, T.~Kim, R.~Koche, Genome-wide maps of chromatin
  state in pluripotent and lineage-committed cells, Nature 448~(7153) (2007)
  553--560.

\bibitem{Siebrasse2008}
J.~P. Siebrasse, R.~Veith, A.~Dobay, H.~Leonhardt, B.~Daneholt, U.~Kubitscheck,
  Discontinuous movement of mrnp particles in nucleoplasmic regions devoid of
  chromatin, Proceedings of the National Academy of Sciences 105~(51) (2008)
  20291--20296.

\bibitem{Pincus1997}
S.~Pincus, R.~E. Kalman, Not all (possibly) ``random'' sequences are created
  equal, Proceedings of the National Academy of Sciences 94~(8) (1997)
  3513--3518.

\bibitem{Rechenberg1982}
J.~Mehra, H.~Rechenberg, The Historical Development of Quantum Theory,
  Springer-Verlag, New York, 1982.

\end{thebibliography}







\end{document}